\documentclass
[superscriptaddress,secnumarabic,amssymb,amsmath,nobibnotes,aps,prd,showkeys,nofootinbib,onecolumn,notitlepage,10pt]{revtex4}%
\usepackage{setspace}
\usepackage{color}
\usepackage{amsmath}
\usepackage{amsfonts}
\usepackage{verbatim}
\usepackage{amssymb}
\usepackage[colorlinks]{hyperref}
\usepackage{graphicx,bm}
\usepackage{amsmath}
\usepackage{amssymb}
\usepackage{graphicx}
\usepackage{textgreek}
\usepackage[caption=false]{subfig}%
\providecommand{\U}[1]{\protect\rule{.1in}{.1in}}

\newcommand{\be}{\begin{equation}}
\newcommand{\ee}{\end{equation}}

\newcommand{\mincir}{\raise
-3.truept\hbox{\rlap{\hbox{$\sim$}}\raise4.truept\hbox{$<$}\ }}
\newcommand{\magcir}{\raise
-3.truept\hbox{\rlap{\hbox{$\sim$}}\raise4.truept\hbox{$>$}\ }}

\ifx\pdfoutput\relax\let\pdfoutput=\undefined\fi
\newcount\msipdfoutput
\ifx\pdfoutput\undefined\else
\ifcase\pdfoutput\else
\ifx\paperwidth\undefined\else
\ifdim\paperheight=0pt\relax\else\pdfpageheight\paperheight\fi
\ifdim\paperwidth=0pt\relax\else\pdfpagewidth\paperwidth\fi
\fi\fi\fi
\begin{document}
\title{Noncoincidence $f(Q)$-Cosmology with Dark Matter Coupled to Gravity}
\author{A. Abebe}
\email{amare.abebe@nithecs.ac.za}
\affiliation{Centre for Space Research, North-West University, Potchefstroom 2520, South Africa}
\affiliation{National Institute for Theoretical and Computational Sciences (NITheCS), South Africa}
\author{P.S. Apostolopoulos}
\email{papostol@ionio.gr}
\affiliation{Mathematical Physics and Computational Statistics Research Laboratory,
Department of Environment, Ionian University, Zakinthos 29100, Greece}
\author{A. Giacomini}
\email{alexgiacomini@uach.cl}
\affiliation{Instituto de Ciencias Fisicas y Matem\`{a}ticas, Universidad Austral de Chile,
Valdivia, Chile}
\author{G. Leon}
\email{genly.leon@ucn.cl}
\affiliation{Departamento de Matem\'{a}ticas, Universidad Cat\`{o}lica del Norte, Avda.
Angamos 0610, Casilla 1280 Antofagasta, Chile}
\affiliation{Institute of Systems Science, Durban University of Technology, Durban 4000,
South Africa}
\author{F. Moncada}
\email{fmoncada@uct.cl}
\affiliation{Departamento de Ciencias Matem\'{a}ticas y F\'{\i}sicas, Universidad Catolica
de Temuco, Temuco, Chile}
\author{A. Paliathanasis}
\email{anpaliat@phys.uoa.gr}
\affiliation{Institute of Systems Science, Durban University of Technology, Durban 4000,
South Africa}
\affiliation{Centre for Space Research, North-West University, Potchefstroom 2520, South Africa}
\affiliation{National Institute for Theoretical and Computational Sciences (NITheCS), South Africa}
\affiliation{Departamento de Matem\'{a}ticas, Universidad Cat\`{o}lica del Norte, Avda.
Angamos 0610, Casilla 1280 Antofagasta, Chile}

\begin{abstract}
We investigate FLRW cosmology in the framework of symmetric teleparallel
$f(Q)$ gravity with a nonminimal coupling between dark matter and the
gravitational field. In the noncoincidence gauge, the field equations admit an
equivalent multi-scalar field representation, which we investigate the
phase-space using the Hubble-normalization approach. We classify all
stationary points for arbitrary function $f(Q)$ and we discuss the physical
properties of the asymptotic solutions. For the power-law theory, we perform a
detailed stability analysis and show that the de Sitter solution is the unique
future attractor, while the matter-dominated point appears as a saddle point.
Moreover, there exist a family of scaling solutions that can be related to
inflationary dynamics. In contrast with uncoupled $f(Q)$ models, the presence
of the coupling introduces a viable matter-dominated era alongside late-time
accelerated expansion. Our study shows that the coupling function plays a
crucial role in cosmological dynamics in $f(Q)$ gravity.

\end{abstract}
\date{\today}
\maketitle

\section{Introduction}

\label{sec1}

Recently, Symmetric Teleparallel General Relativity (STEGR)
\cite{Nester:1998mp} and its generalizations have attracted considerable
attention in cosmology as promising frameworks for describing the dynamical
structure of the universe.

In General Relativity (GR) the fundamental object associated with the
gravitational field is the Levi-Civita connection. In contrast, in STEGR the
corresponding role is played by the nonmetricity tensor, defined through a
symmetric and flat connection \cite{revh}. Although GR and STEGR are
equivalent, this equivalence breaks down when scalar fields or nonlinear
extensions of the geometric scalars are introduced into the gravitational
Action \cite{Koivisto2,Koivisto3,Baha1,sc1,sc3,mf1,mf2,mf3,Shabani:2025qxn}.
For instance, the $f\left(  R\right)  $ and $f\left(  Q\right)  $ theories of
gravity are equivalent only when the functional dependence is linear, that is,
the theories are equivalent with GR or STEGR respectively. This simple
observation has been omitted in a series of studies as discussed recently in
\cite{tel}.

The choice of connection within the symmetric teleparallel theory is not
unique, and selecting an appropriate one is crucial for ensuring the physical
viability of the model and for defining the gravitational theory itself. As
discussed in detail in \cite{dd2}, a number of studies in the literature have
employed nonphysical connections in the investigation of static spherically
symmetric spacetimes. On the other hand, in \cite{dd4} it was shown that in a
Kantowski-Sachs geometry tilted fluid components can be supported, in contrast
with GR where this feature is not possible for such an anisotropic background.

In modern cosmology, where spacetime is usually modeled by the isotropic and
homogeneous Friedmann-Lema\^{\i}tre-Robertson-Walker (FLRW) geometry, there
exist four possible connections consistent with being symmetric, flat, and
compatible with the isometries of the background \cite{Hohmann,fq4}. Three of
these correspond to the spatially flat case, while the fourth applies to the
nonzero-curvature case. The associated degrees of freedom of $f\left(
Q\right)  $ theory have been investigated in detail in the literature. In
particular, the field equations admit a minisuperspace description \cite{an3}.
It was found that when the connection is defined in the noncoincidence gauge,
additional dynamical degrees of freedom appear, which can be interpreted as
scalar fields \cite{an2}. On the other hand, when the coincidence gauge is
adopted, the resulting cosmological equations coincide with those of $f\left(
T\right)  $ teleparallel gravity \cite{tel1}. Although torsion teleparallel
theory of gravities, can violate the local Lorentz symmetry in some
extensions, for details we refer to the discussion in \cite{lo1,lo2,lo3}, this
is not the case of symmetric teleparallel theory.

Within the coincidence gauge, observational tests of $f\left(  Q\right)  $
theory have been presented in \cite{ob1,ob2,ob3}. More recently, the issues of
structure growth and the $H_{0}$ and $S_{8}$ tensions were examined in
\cite{ob4,ob5}. The effects of adopting the noncoincidence connection on the
description of late-time acceleration were analyzed in \cite{an4}. In
particular, by applying the latest Baryonic Acoustic Oscillation (BAO) data
from the DESI DR2 release \cite{des4,des5,des6}, it was found that this
cosmological model challenges $\Lambda$CDM and related theories formulated in
the coincidence gauge.

In \cite{ppr1,ppr2} it was shown that $f\left(  Q\right)  $ gravity suffers
from strong coupling or the appearance of ghosts at the level of cosmological
perturbations. Nevertheless, these problems can be avoided when the matter
sector is coupled to the gravitational field directly in the Action Integral.
This type of interaction is the focus of the present work, where we restrict
attention to the background dynamics. Specifically, we perform a detailed
phase-space analysis of the cosmological field equations in the presence of a
coupling function. The stationary points are identified and used to
reconstruct the possible cosmological histories supported by the theory.
Cosmological models with an interacting dark sector have drawn the attention
recently because they can describe the recent cosmological data
\cite{dt1,dt2,dt3,dt4}.\ It was found that the late-time universe indicate a
strong interaction coupling between the dark energy and the dark matter
\cite{dt5,dt6,dt7}. In $f\left(  Q\right)  $ gravity, dark energy has a
geometric origin, thus in order to introduce the interaction with the dark
matter component, we consider that the Lagrangian function of the latter is
coupled with a nonlinear function to the nonmetricity scalar $Q$. 

Because of the nonlinearity of the gravitational field equations, phase-space
methods provide a powerful tool for understanding the evolution of physical
properties in cosmological models \cite{cop1}. Such analyses have been widely
applied in various modified gravity theories
\cite{ex1,ex2,ex3,ex4,ex5,ex6,ex7,ex8,ex9,ex10}, yielding valuable information
on constraints for free parameters and on the behavior of the initial value
problem \cite{ex11,ex12}. A detailed analysis of the phase-space of $f\left(
Q\right)  $ cosmology was carried out in \cite{an1}, where it was shown that
different choices of connection lead to distinct sets of field equations and
hence to different cosmic evolutions. For additional work on phase-space
analyses in modified STEGR, we refer the reader to
\cite{fq5,fq6,fq7,anc1,anc2} and references therein.

The purpose of this study is to investigate the impact of a coupling function
between dark matter and gravity on the background dynamics and to assess the
physical viability of the theory. The structure of the paper is as follows.

In Section \ref{sec2}, we briefly review STEGR and its generalization, the
$f\left(  Q\right)  $-theory. The cosmological model under consideration is
introduced in Section \ref{sec3}, where we examine a coupling between the
matter source and the gravitational field. The coupling function is defined in
such a way that, in the scalar-field representation, the field equations take
their simplest form. Our analysis is carried out in the context of the
noncoincident connection, where the connection introduces nontrivial dynamical
degrees of freedom within the field equations. The corresponding cosmological
field equations are equivalent to a multi-scalar field gravitational theory.

The existence of cosmological solutions of particular interest is investigated
in Section \ref{sec4}. Specifically, we focus on the power-law solution
describing the matter-dominated era and on the de Sitter universe. In Section
\ref{sec5}, we present a detailed dynamical analysis of the cosmological field
equations and explore the impact of the matter-gravity coupling on the global
cosmic history. We determine the stationary points for an arbirtary function
$f\left(  Q\right)  $, nevertheless we investigate the stability properties
for the special case of the power-law $f\left(  Q\right)  \simeq Q^{\frac
{n}{n-1}}$ model. Finally, in Section \ref{sec6}, we summarize our findings
and present our conclusions.

\section{Symmetric Teleparallel Gravity}

\label{sec2}

Consider a four-dimensional manifold $M^{4}$, equipped with a metric tensor
$g_{\mu\nu}$ and a generic affine connection $\Gamma_{\mu\nu}^{\kappa}$, which
defines the covariant derivative $\nabla_{\lambda}$. The generic connection
can be decomposed into three components \cite{Eisenhart}.

The Levi-Civita connection $\mathring{\Gamma}_{\mu\nu}^{\kappa}$, defined by
the metric tensor
\begin{equation}
\mathring{\Gamma}_{\mu\nu}^{\kappa}=\frac{1}{2}g^{\kappa\lambda}\left(
g_{\mu\lambda,\nu}+g_{\lambda\nu,\mu}-g_{\mu\nu,\lambda}\right)  ,
\end{equation}
the torsion tensor $\mathrm{T}_{\mu\nu}^{\kappa}$, defined as the
antisymmetric part of the connection
\begin{equation}
\mathrm{T}_{\mu\nu}^{\kappa}=\Gamma_{\mu\nu}^{\kappa}-\Gamma_{\nu\mu}^{\kappa
},
\end{equation}
and the nonmetricity tensor $Q_{\lambda\mu\nu}$, which measures the failure of
the connection to preserve the metric
\begin{equation}
Q_{\lambda\mu\nu}=\partial_{\lambda}g_{\mu\nu}-\Gamma_{\lambda\mu}^{\sigma
}g_{\sigma\nu}-\Gamma_{\lambda\nu}^{\sigma}g_{\mu\sigma}.
\end{equation}

Consequently, the generic connection $\Gamma_{\mu\nu}^{\kappa}$ is expressed
as follows \cite{Eisenhart}:
\begin{equation}
\Gamma_{\mu\nu}^{\kappa}=\mathring{\Gamma}_{\mu\nu}^{\kappa}+\mathrm{T}%
_{\mu\nu}^{\kappa}+Q_{\;\mu\nu}^{\kappa}.
\end{equation}
Each component of the connection defines a fundamental tensor that
characterizes the geometric structure of the manifold $M^{4}$.

In GR, the gravitational field is described solely by the Levi-Civita
connection, i.e., $\Gamma_{\mu\nu}^{\kappa}=\mathring{\Gamma}_{\mu\nu}%
^{\kappa}$, and the dynamics are given by the Riemann curvature tensor:
\begin{equation}
R_{\;\lambda\mu\nu}^{\kappa}=\partial_{\mu}\Gamma_{\lambda\nu}^{\kappa
}-\partial_{\nu}\Gamma_{\lambda\mu}^{\kappa}+\Gamma_{\lambda\nu}^{\sigma
}\Gamma_{\mu\sigma}^{\kappa}-\Gamma_{\lambda\mu}^{\sigma}\Gamma_{\nu\sigma
}^{\kappa}.
\end{equation}
In contrast, in TEGR, the connection is taken to be purely antisymmetric,
i.e., $\Gamma_{\mu\nu}^{\kappa}=\mathrm{T}_{\mu\nu}^{\kappa}$, and the
gravitational field is described by the torsion tensor. Finally, in STEGR, the
{connection is assumed to be flat $R_{\;\lambda\mu\nu}^{\kappa}=0$} and
symmetric $\mathrm{T}_{\mu\nu}^{\kappa}=0$, implying that only the
nonmetricity tensor contributes to the gravitational field $\Gamma_{\mu\nu
}^{\kappa}=Q_{\;\mu\nu}^{\kappa}.$

The analogue of the Einstein-Hilbert action in STEGR is given by
\cite{Nester:1998mp}%
\begin{equation}
S_{Q}=\int d^{4}x\sqrt{-g}Q, \label{ac1}%
\end{equation}
where the nonmetricity scalar $Q$ is defined as
\[
Q=Q_{\lambda\mu\nu}P^{\lambda\mu\nu},
\]
and $P_{\;\mu\nu}^{\lambda}$ is the the conjugate tensor with definition
\begin{equation}
P_{\;\mu\nu}^{\lambda}=-\frac{1}{4}Q_{\;\mu\nu}^{\lambda}+\frac{1}{2}%
Q_{\;(\mu\nu)}^{\lambda}+\frac{1}{4}\left(  Q^{\lambda}-\bar{Q}^{\lambda
}\right)  g_{\mu\nu}-\frac{1}{4}\delta_{(\mu}^{\lambda}Q_{\nu)}.
\end{equation}
Parentheses denote symmetrization: $A_{(\mu\nu)}=\frac{1}{2}(A_{\mu\nu}%
+A_{\nu\mu})$; $Q_{\mu}=Q_{\mu\nu}^{\;\;\;\nu}$, $\bar{Q}_{\mu}=Q_{\;\mu\nu
}^{\nu}$, and $\delta_{\nu}^{\mu}$ is the Kronecker delta.

\subsection{$f\left(  Q\right)  $-gravity}

Introducing nonlinear functions of the nonmetricity scalar $Q$ in the action
(\ref{ac1}) leads to the family of $f(Q)$-gravity theories, with the action
\cite{Koivisto2,Koivisto3}
\begin{equation}
S_{f(Q)}=\int d^{4}x\sqrt{-g}f(Q). \label{ac2}%
\end{equation}

In the absence of matter, variation of the action (\ref{ac2}) with respect to
the metric yields the field equations \cite{revh}
\begin{equation}
\frac{2}{\sqrt{-g}}\nabla_{\lambda}\left(  \sqrt{-g}f_{\;\mu\nu}%
^{\prime\lambda}\right)  -\frac{f(Q)}{2}g_{\mu\nu}+f^{\prime}(Q)\left(
P_{\mu\rho\sigma}Q_{\nu}^{\;\rho\sigma}-2Q_{\rho\sigma\mu}P_{\;\;\;\nu}%
^{\rho\sigma}\right)  =0,
\end{equation}
where $f^{\prime}(Q)=\frac{df}{dQ}$. Variation with respect to the connection
yields the equation of motion%
\begin{equation}
\nabla_{\mu}\nabla_{\nu}\left(  \sqrt{-g}f^{\prime}\left(  Q\right)
P_{\;\;\;\sigma}^{\mu\nu}\right)  =0. \label{kl.01}%
\end{equation}

This equation is trivially satisfied in the so-called coincidence gauge, where
the connection vanishes. However, as discussed in various gravitational
models, $f(Q)$-gravity recovers the GR limit only when a non-coincidence gauge
is employed \cite{anc3}. Therefore, equation (\ref{kl.01}) plays a crucial
role in the gravitational dynamics \cite{ndim1}. For a recent discussion we
refer the reader to \cite{dial1}.

\section{Matter Coupled to Gravity}

\label{sec3}

On cosmological scales, the universe is assumed to be isotropic and
homogeneous, and is described by the FLRW metric with the line element
\begin{equation}
ds^{2}=-N^{2}(t)\,dt^{2}+a^{2}(t)\left(  dx^{2}+dy^{2}+dz^{2}\right)  ,
\label{kd.02}%
\end{equation}
where $N(t)$ is the lapse function and $a(t)$ is the scale factor. For this
line element there are three families of connections which lead to different
cosmological models \cite{Hohmann,fq4}. In this work, we consider the
nontrivial affine connection with the following nonzero components:
\begin{equation}
\Gamma_{tt}^{t}=\frac{\ddot{\psi}(t)}{\dot{\psi}(t)}+\dot{\psi}(t),
\end{equation}%
\[
\Gamma_{tr}^{r}=\Gamma_{rt}^{r}=\Gamma_{t\theta}^{\theta}=\Gamma_{\theta
t}^{\theta}=\Gamma_{t\varphi}^{\varphi}=\Gamma_{\varphi t}^{\varphi}=\dot
{\psi}(t).
\]

The nonmetricity scalar $Q$ is calculated as
\begin{equation}
Q=-6H^{2}+\frac{3\dot{\psi}}{N}\left(  3H-\frac{\dot{N}}{N^{2}}\right)
+\frac{3\ddot{\psi}}{N^{2}}, \label{kd.03}%
\end{equation}
where $H=\frac{1}{N}\frac{\dot{a}}{a}$ is the Hubble parameter.

It has been shown that this connection leads to cosmological dynamics with de
Sitter behavior as an attractor, even without introducing a cosmological
constant \cite{an1,dut1}, unlike the case with the coincidence connection.
Moreover, inflationary solutions can be realized within this framework, as the
theory is dynamically equivalent to a two-scalar field quintom-like
gravitational model \cite{an2}. Recently, in \cite{an4} it was found that the
geometrodynamical degrees of freedom introduced by this noncoincident
connection can describe the late-time acceleration of the universe. The
$\Lambda$CDM limit within the $f\left(  Q\right)  $-gravity for this
connection was examined before in \cite{saik}. On the other hand, this
connection has been applied to investigate the evolution of the anisotropies
in a homogeneous Bianchi I \ universe \cite{saik2}.

In this work, we introduce the matter Lagrangian $L_{m}=\rho_{m}$,
representing an ideal pressureless fluid that models dark matter. The coupling
to gravity is described by the action
\begin{equation}
S_{f(Q)}=\int d^{4}x\sqrt{-g}\left[  f(Q)-\alpha f^{\prime}(Q)L_{m}\right]
,\label{kd.04}%
\end{equation}
such that in the limit $f(Q)\sim Q$, general relativity is recovered. Such
interacting models have been examined before in other modified theories of
gravity, see for instance \cite{fr1,fr2,fr3}. \ In our consideration the
interacting function has been introduced in a way such that to introduce the
minimum nonlinear terms in the cosmological field equations. The form of the
coupling function  has been considered such that to have the minimum number of
degrees of freedom introduced by the coupling function and have similarities
with other theories of gravity such as the Weyl Integrable Spacetime or the
Chameleon Mechanism. 

\bigskip

Assuming dark matter behaves as a pressureless fluid, we take $L_{m}=\rho
_{m0}a^{-3}$. Introducing a Lagrange multiplier $\lambda$, the action becomes
\begin{equation}
S_{f(Q)}=\int d^{4}x\sqrt{-g}\left\{  f(Q)+\lambda\left[  Q-\left(
-6H^{2}+\frac{3\dot{\psi}}{N}\left[  3H-\frac{\dot{N}}{N^{2}}\right]
+\frac{3\ddot{\psi}}{N^{2}}\right)  \right]  -\alpha f^{\prime}(Q)\rho
_{m0}a^{-3}\right\}  ,
\end{equation}
where $\lambda=-f^{\prime}(Q)$.

After integrating by parts, the point-like action becomes
\begin{equation}
S_{f(Q)}=\int L_{f(Q)}(N,a,\dot{a},\phi,\dot{\phi},\psi,\dot{\psi})\,dt,
\end{equation}
with the Lagrangian \cite{an3}
\begin{equation}
L_{f(Q)}=-\frac{3}{N}\phi a\dot{a}^{2}-\frac{3}{2N}a^{3}\dot{\phi}\dot{\psi
}+Na^{3}V(\phi)-\beta N\phi, \label{lan.01}%
\end{equation}
where $\beta=\alpha\rho_{m0}$, $\phi=f^{\prime}(Q)$, and the potential is
defined as $V(\phi)=f(Q)-Qf^{\prime}(Q)$. The inverse relation is given by
\begin{equation}
f(Q)=V(\phi(Q))-\phi(Q)V_{,\phi}(Q). \label{cc1}%
\end{equation}
Within the minisuperspace description it is easy to see that under a conformal
transformation, the coupling function between the gravitational field and the
matter is eliminated.

The cosmological field equations follow from the variation of the latter
point-like Lagrangian with respect to the dynamical variables $\left\{
N,a,\phi,\psi\right\}  $. Setting $N=1$ we calculate the equations
\begin{align}
3\phi H^{2}+\frac{3}{2}\dot{\phi}\dot{\psi}+V(\phi)-\beta\phi a^{-3}  &
=0,\label{lan.02}\\
\phi(2\dot{H}+3H^{2})+2H\dot{\phi}-\frac{3}{2}\dot{\phi}\dot{\psi}+V(\phi)  &
=0,\\
\ddot{\phi}+3H\dot{\phi}  &  =0,\\
\ddot{\psi}-2H^{2}+3H\dot{\psi}+\frac{2}{3}V_{,\phi}  &  =-\frac{2}{3}\beta
a^{-3}.
\end{align}

Solving for the higher-order derivatives we find%
\begin{align}
\dot{H}  &  =-\frac{H\dot{\phi}}{\phi}-\frac{3}{2}H^{2}-\frac{V(\phi)}{2\phi
}+\frac{3\dot{\psi}\dot{\phi}}{4\phi},\\
\ddot{\phi}  &  =-3H\dot{\phi},\\
\ddot{\psi}  &  =-\frac{2\beta}{3a^{3}}-3H\dot{\psi}+2H^{2}-\frac{2}%
{3}V^{\prime}(\phi).
\end{align}

The modified Friedmann equations can be written in the equivalent form%
\begin{align}
3H^{2}  &  =\frac{1}{\phi}\left(  \rho_{f\left(  Q\right)  }+\rho_{m}\right)
,\\
2\dot{H}+3H^{2}  &  =-\frac{1}{\phi}p_{f\left(  Q\right)  },
\end{align}
where
\begin{align}
\rho_{f\left(  Q\right)  }  &  =-\frac{3}{2}\dot{\phi}\dot{\psi}-V\left(
\phi\right)  ,\\
p_{f\left(  Q\right)  }  &  =-\frac{3}{2}\dot{\phi}\dot{\psi}+2H\dot{\phi
}+V\left(  \phi\right)  ,
\end{align}
are the geometric fluid components related to the $f\left(  Q\right)
$-gravity, and $\rho_{m}=\beta\phi a^{-3}$, is the energy density for the
matter source, which satisfies the conservation law%
\begin{equation}
\dot{\rho}_{m}+3H\rho_{m}=\rho_{m}\left(  \ln\phi\right)  ^{\cdot}\text{.}%
\end{equation}

In the following, we investigate the impact of the coupling parameter $\beta$
on the cosmological dynamics of $f(Q)$-gravity.

\section{Exact Solutions}

\label{sec4}

We examine the existence of two exact cosmological solutions of special
interests, the power-law solution, $a\left(  t\right)  =a_{0}t^{p}$, with
Hubble function $H\left(  t\right)  =\frac{p}{t}$, and the de Sitter solution
$a\left(  t\right)  =a_{0}e^{H_{0}t}$ with constant Hubble function $H\left(
t\right)  =H_{0}$. The power-law solution can describe a radiation epoch, for
$p=\frac{1}{2}$, or the matter dominated era, for $p=\frac{2}{3}$. This
analysis will reveal important information about the behaviour of the
$f\left(  Q\right)  $-function regarding the existence of these main eras.

\subsection{Power-law Solution}

Substituting $H(t)=\frac{p}{t}$ into the cosmological field equations, we find
that the system admits the following solution:
\begin{align}
\phi(t) &  =\phi_{1}t^{1-3p}+\phi_{0},\\
\dot{\psi}(t) &  =-\frac{\phi_{0}\left(  2pt^{-2+3p}-\beta\right)  +\phi
_{1}\left(  6p^{2}t^{-1}-\beta t^{1-3p}\right)  }{3\phi_{1}(1-3p)},\\
V(t) &  =(1-3p)p\phi_{0}t^{-2}+\frac{\beta}{2}t^{-6p}\left(  \phi_{0}%
t^{3p}+\phi_{1}t\right)  .
\end{align}

The corresponding scalar potential $V(\phi)$ can be reconstructed as
\begin{align}
V(\phi) &  =(3p-1)\phi_{0}\left(  \phi_{1}(\phi-\phi_{0})\right)  ^{\frac
{2}{3p-1}}\nonumber\\
&  ~~~~~~-\frac{\beta}{2}\left(  \phi_{1}(\phi-\phi_{0})\right)  ^{1+\frac
{1}{3p-1}}\phi.
\end{align}
Indeed, in the absence of the interacting term, that is, $\beta=0$, the latter
potential leads to the power-law function for the $f\left(  Q\right)  $-model
as discussed before in \cite{ndim1}. 

\subsection{de Sitter Solution}

For the de Sitter case, where $H(t)=H_{0}$ is constant, the field equations
admit the solution:
\begin{align}
\phi(t)  &  =\phi_{1}e^{-3H_{0}t}+\phi_{0},\\
\dot{\psi}(t)  &  =\frac{2}{3}H_{0}-\frac{\beta}{9H_{0}}\left(  e^{-3H_{0}%
t}+\frac{\phi_{0}}{\phi_{1}}\right)  ,\\
V(t)  &  =-3H_{0}^{2}\phi_{0}-\frac{\beta}{2}e^{-6H_{0}t}\left(  e^{3H_{0}%
t}\phi_{0}+\phi_{1}\right)  .
\end{align}

The scalar potential $V(\phi)$ takes the form%
\begin{equation}
V(\phi)=-3H_{0}^{2}\phi_{0}+\frac{\beta}{2\phi_{1}}(\phi-\phi_{0})\phi.
\end{equation}

Consequently, the closed-form expression for the function $f(Q)$ is
\begin{equation}
f(Q)=f_{1}\frac{\phi_{1}}{\sqrt{2\beta}}Q-\frac{\phi_{1}}{2\beta}Q^{2}%
+\frac{\phi_{1}f_{1}^{2}}{4}-3H_{0}^{2}\phi_{0}.
\end{equation}

Therefore, only when $\phi_{1}=0\,,$ that is, $\phi\left(  t\right)  =\phi
_{0}$, the interacting term can be omitted, i.e. $\beta=0$, which leads to the
case $f\left(  Q\right)  =-3H_{0}^{2}\phi_{0}$, that is $Q$ is a constant
scalar \cite{ndim1}.

In the following, we perform a detailed analysis for the phase-space for the
cosmological field equations.\ Such analysis will provide us with important
information about the nature of the interacting term in the cosmic evolution.

\section{Cosmological Dynamics}

\label{sec5}

A systematic exploration of the dynamical structure of scalar--tensor models
offers profound insights into their cosmological behavior. By recasting the
field equations into an autonomous system with suitable dimensionless
variables, one gains direct access to the global phase--space geometry. This
framework makes it possible to identify fixed points such as attractors,
repellers, and saddle points that determine the asymptotic evolution of the universe.

\subsection{Dimensionless Variables and Dynamical System}

We work within the $H$-normalization approach and we define the following
dimensionless variables
\begin{equation}%
\begin{pmatrix}
\Omega_{m}\\
\Omega_{V}\\
x_{\phi}\\
x_{\psi}\\
\lambda
\end{pmatrix}
=%
\begin{pmatrix}
\frac{1}{3a^{3}H^{2}}\\
\frac{V(\phi)}{3H^{2}\phi}\\
\frac{\dot{\phi}}{H\phi}\\
\frac{\dot{\psi}}{2H}\\
\frac{\phi V^{\prime}(\phi)}{V(\phi)}%
\end{pmatrix}
\end{equation}

These variables quantify the normalized energy densities and field velocities.
$\Omega_{m}$ describes the matter contribution,~$\Omega_{V}$ the potential
energy contribution,~$x_{\phi}$ is the normalized velocity of the scalar field
$\phi$ and $x_{\psi}$ defines the normalized velocity of the scalar field
$\psi$. \
\begin{align}
\lambda &  =\frac{\phi V^{\prime}(\phi)}{V(\phi)},\\
g(\lambda)  &  =\frac{\phi^{2}V^{\prime\prime}(\phi)}{V(\phi)}-\frac{\phi
^{2}V^{\prime2}}{V(\phi)^{2}}+\frac{\phi V^{\prime}(\phi)}{V(\phi)}.
\end{align}

The first modified Friedmann equation (\ref{lan.02}) is expressed as the
algebraic constraint
\begin{equation}
x_{\psi}x_{\phi}-\beta\Omega_{m}+\Omega_{V}+1=0.
\end{equation}
The latter constraint allows to reduce by one the dimension of the dynamical system.

Using the e-folding number $N=\ln a$ as the time variable, the autonomous
system becomes
\begin{equation}
\frac{d}{dN}%
\begin{pmatrix}
\Omega_{V}\\
x_{\phi}\\
x_{\psi}\\
\lambda
\end{pmatrix}
=%
\begin{pmatrix}
\Omega_{V}\left(  \lambda x_{\phi}-3x_{\psi}x_{\phi}+x_{\phi}+3\Omega
_{V}+3\right) \\
-\frac{3}{2}x_{\phi}\left(  x_{\psi}x_{\phi}-\Omega_{V}+1\right) \\
-\lambda\Omega_{V}-\frac{3x_{\psi}^{2}x_{\phi}}{2}+\frac{1}{2}x_{\psi}\left(
4x_{\phi}+3\Omega_{V}-3\right)  +\Omega_{V}+2\\
x_{\phi}g(\lambda)
\end{pmatrix}
. \label{systA}%
\end{equation}
Function $g(\lambda)$ is defined as%
\begin{equation}
g(\lambda\left(  \phi\right)  )=\frac{\phi^{2}V^{\prime\prime}(\phi)}{V(\phi
)}-\frac{\phi^{2}V^{\prime2}}{V(\phi)^{2}}+\frac{\phi V^{\prime}(\phi)}%
{V(\phi)}.
\end{equation}

The evolution equation for $\Omega_{m}$ is decoupled as follows
\begin{equation}
\frac{d}{dN}\Omega_{m}=-\left(  (3x_{\psi}-2)x_{\phi}-3\Omega_{V}\right)
\left(  x_{\psi}x_{\phi}+\Omega_{V}+1\right)  ,
\end{equation}
which confirms that $\Omega_{m}$ is evolving once the other variables are known.

Furthermore, in terms of the new variables, the effective equation of state
parameter $w_{eff}=-1-\frac{2}{3}\frac{\dot{H}}{H^{2}}$ reads%
\begin{equation}
w_{eff}=x_{\phi}\left(  \frac{2}{3}-x_{\psi}\right)  +\Omega_{V}\text{.}%
\end{equation}

\subsection{Equilibrium Points and Cosmological Interpretation}

The stationary points of the system \eqref{systA} correspond to cosmological
regimes where the dynamical variables remain constant. We calculate the
stationary points for a generic function $g\left(  \lambda\right)  $. Recall
that each function $g\left(  \lambda\right)  $ corresponds to a given scalar
field potential. Indeed, for $g\left(  \lambda\right)  =0$, we calculate that
$V(\phi)=V_{0}\phi^{\lambda}$. On the other hand for $g\left(  \lambda\right)
=-\lambda^{2}$, the corresponding scalar field potential is derived
$V(\phi)=V_{0}+V_{1}\ln(\phi)$; while for $g(\lambda)=\lambda_{0}^{2}%
-\lambda^{2}$ we calculate $V(\phi)=V_{0}\phi^{-\lambda_{0}}+V_{1}%
\phi^{\lambda_{0}}$.

Furthermore, using the Clairaut equation (\ref{cc1}) the corresponding
$f\left(  Q\right)  $ function can be reconstructed. We shall focus in the
power-law potential $V(\phi)=V_{0}\phi^{\lambda}$, from where determine the
power-law theory $f\left(  Q\right)  \simeq Q^{\frac{\lambda}{\lambda-1}}$.

The stationary points for the four dimensional dynamical system are presented
in Table \ref{point1}. \ Each stationary is characterized by its coordinates
and the corresponding value of $\lambda$, which encodes the steepness of the
potential. The condition $\lambda_{\ast}\in g^{-1}(0)$ ensures that the
potential curvature vanishes at the fixed point, allowing $\lambda$ to remain constant.

The asymptotic solutions at the stationary points can be categorized into the
following three families.

(I)\ The Potential dominated solutions (Points A and D) where the scalar field
$\phi$ is frozen and the potential energy dominates. The stationary points
describe de Sitter expansion because $w_{eff}=-1$.

(II) The Kinetic dominated solution (Point B) Scalar field $\phi$ is frozen,
but $\psi$ evolves dynamically. It describes a matter dominated solution,
$\Omega_{m}=\frac{1}{\beta}$ and $w_{eff}=0$. The requirement $\Omega_{m}%
\geq0$, leads to the constraint $\beta>0$.

(III)\ The Scaling solutions (Points C and E) the scalar fields evolve
proportionally to the expansion rate, allowing for tracking behavior or
late-time acceleration. We calculate that there is not any contribution from
the matter to the cosmic fluid, i.e. $\Omega_{m}=0$. Moreover, it follows
$w_{eff}\left(  C\right)  =1-\frac{2}{3x_{\psi}}$ $\ $and $w_{eff}\left(
E\right)  =-1-\frac{4}{1+\lambda_{\ast}}$. Thus acceleration is occurred when
$0<x_{\psi}<\frac{1}{2}$, or $-1<\lambda_{\ast}<2$.%

\begin{table}[tbp] \centering
\caption{Stationary points for the cosmological field equations.}%
\begin{tabular}
[c]{cccccccp{7.5cm}}\hline\hline
\textbf{Point} & $\mathbf{\Omega}_{V}$ & $\mathbf{x}_{\phi}$ & $\mathbf{x}%
_{\psi}$ & $\mathbf{\lambda}$ & $\mathbf{\Omega}_{m}$ & $\mathbf{w}_{eff}$ &
\multicolumn{1}{c}{\textbf{Interpretation}}\\\hline
$A$ & $-1$ & $0$ & $x_{\psi}$ & $3x_{\psi}-1$ & $0$ & $-1$ & Scalar field
$\phi$ is static, while $\psi$ evolves. The potential energy dominates with
$\Omega_{V}=-1$, suggesting a vacuum-like phase with non-standard energy
density. This may reflect modified gravity effects or a noncanonical
potential.\\
$B$ & $0$ & $0$ & $\frac{4}{3}$ & arbitrary & $-\frac{1}{\beta}$ & $0$ &
Scalar field $\phi$ is frozen, but $\psi$ evolves dynamically. The potential
energy vanishes, and the dynamics are driven by the kinetic energy of $\psi$.
This resembles a kinetic-dominated regime, relevant in early universe
scenarios.\\
$C(\lambda_{\ast})$ & $0$ & $-\frac{1}{x_{\psi}}$ & $x_{\psi}$ &
$\lambda_{\ast}\in g^{-1}(0)$ & $0$ & $1-\frac{2}{3x_{\psi}}$ & A scaling
solution where both scalar fields evolve in proportion to the Hubble
expansion. The potential energy is negligible, and the system exhibits
tracking behavior. This point can act as an attractor and help resolve the
coincidence problem.\\
$D(\lambda_{\ast})$ & $-1$ & $0$ & $\frac{1}{3}(\lambda_{\ast}+1)$ &
$\lambda_{\ast}\in g^{-1}(0)$ & $0$ & $-1$ & Scalar field $\phi$ is static,
and the potential dominates. The evolution is driven by $\psi$, and the
configuration may mimic a dark energy-like phase. Stability depends on the
curvature of the potential via $g(\lambda_{*})$.\\
$E(\lambda_{\ast})$ & $0$ & $-\frac{6}{\lambda_{\ast}+1}$ & $\frac{1}%
{6}(\lambda_{\ast}+1)$ & $\lambda_{\ast}\in g^{-1}(0)$ & $0$ & $1-\frac
{4}{1+\lambda_{\ast}}$ & An enhanced scaling regime where both scalar fields
evolve dynamically. The potential energy vanishes, but kinetic terms
contribute significantly. This configuration can support late-time accelerated
expansion if $\lambda_{*}>0$ and $g^{\prime}(\lambda_{*})>0$.\\\hline\hline
\end{tabular}
\label{point1}%
\end{table}%

\subsection{Linear Stability Analysis}

In order to investigate the stability of each stationary point, we linearize
the system around $\mathbf{x}=(\Omega_{V},x_{\phi},x_{\psi},\lambda)$ and
compute the Jacobian matrix $J=\frac{\partial\mathbf{f}}{\partial\mathbf{x}}$.

The Jacobian matrix of the linearized system reads%
\[
J=%
\begin{pmatrix}
3\left(  1+2\Omega_{V}\right)  +x_{\phi}\left(  1-3x_{\psi}+\lambda\right)  &
\left(  1-3x_{\psi}+\lambda\right)  \Omega_{V} & -3x_{\phi}\Omega_{V} &
x_{\phi}\Omega_{V}\\
\frac{3}{2}x_{\phi} & \frac{3}{2}\left(  \Omega_{V}-1-2x_{\phi}x_{\psi}\right)
& -\frac{3}{2}x_{\phi}^{2} & 0\\
1+\frac{3}{2}x_{\psi}-\lambda & \frac{\left(  4-3x_{\psi}\right)  x_{\psi}}{2}
& \frac{\left(  4x_{\phi}-3\left(  \left(  1+\Omega_{V}\right)  -2x_{\phi
}x_{\psi}\right)  \right)  }{2} & -\Omega_{V}\\
0 & g\left(  \lambda\right)  & 0 & x_{\phi}g_{,\lambda}%
\end{pmatrix}
\text{.}%
\]
The eigenvalues of $J$ at the stationary point determine the nature of the
solution. Specifically, if all eigenvalues with negative real parts imply a
stable attractor. If all positive real parts indicate an unstable repeller. On
the other hand, mixed signs on the real parts of the eigenvalues correspond to
a saddle point, while zero real parts signal a non-hyperbolic point, requiring
center manifold or higher-order analysis.

The definition of the $g\left(  \lambda\right)  $ function is essential for
the stability analysis. Thus, for this in the following we consider the case
$g\left(  \lambda\right)  =0$, where the dimension of the dynamical system is
reduced by one.

\subsubsection{Power-law potential}

For the power-law potential, the stationary points $B$, $C\,$\ and $D$ exist
for arbitrary value of parameter $\lambda$. Recall that in this case $\lambda$
is a constant, and point $E$ becomes a special case of point $C$. The reduced
Jacobian matrix is calculated
\[
J=%
\begin{pmatrix}
3\left(  1+2\Omega_{V}\right)  +x_{\phi}\left(  1-3x_{\psi}+\lambda\right)  &
\left(  1-3x_{\psi}+\lambda\right)  \Omega_{V} & -3x_{\phi}\Omega_{V}\\
\frac{3}{2}x_{\phi} & \frac{3}{2}\left(  \Omega_{V}-1-2x_{\phi}x_{\psi}\right)
& -\frac{3}{2}x_{\phi}^{2}\\
1+\frac{3}{2}x_{\psi}-\lambda & \frac{\left(  4-3x_{\psi}\right)  x_{\psi}}{2}
& \frac{\left(  4x_{\phi}-3\left(  \left(  1+\Omega_{V}\right)  -2x_{\phi
}x_{\psi}\right)  \right)  }{2}%
\end{pmatrix}
.
\]
For the stationary point $B$, the corresponding eigenvalues are $\left\{
3,-\frac{3}{2},-\frac{3}{2}\right\}  $, from where we infer that the
stationary point is a saddle point. Moreover, for point $D$, the corresponding
eigenvalues are $\left\{  -3,-3,-3\right\}  $, indicate that the de Sitter
solution is always an attractor. Finally, for the stationary point $C$ we
calculate the eigenvalues $\left\{  0,3-\frac{2}{3x_{\psi}},\frac{6x_{\psi
}-\lambda-1}{x_{\psi}}\right\}  $, where for $\left\{  \lambda\leq3,x_{\psi
}>\frac{2}{3}\right\}  $ or $\left\{  1+\lambda>0,x_{\psi}<0\right\}  $ or
$\left\{  \lambda>3,x_{\psi}>\frac{1+\lambda}{6}\right\}  $ or $\left\{
\lambda\leq-1,x_{\psi}<1+\lambda\right\}  $, the point is a source. Otherwise
for $\left\{  x_{\psi}<\frac{2}{3},\frac{6x_{\psi}-\lambda-1}{x_{\psi}%
}<0\right\}  $ the stationary point is an attractor.

\begin{figure}[ptbh]
\centering\includegraphics[width=1\textwidth]{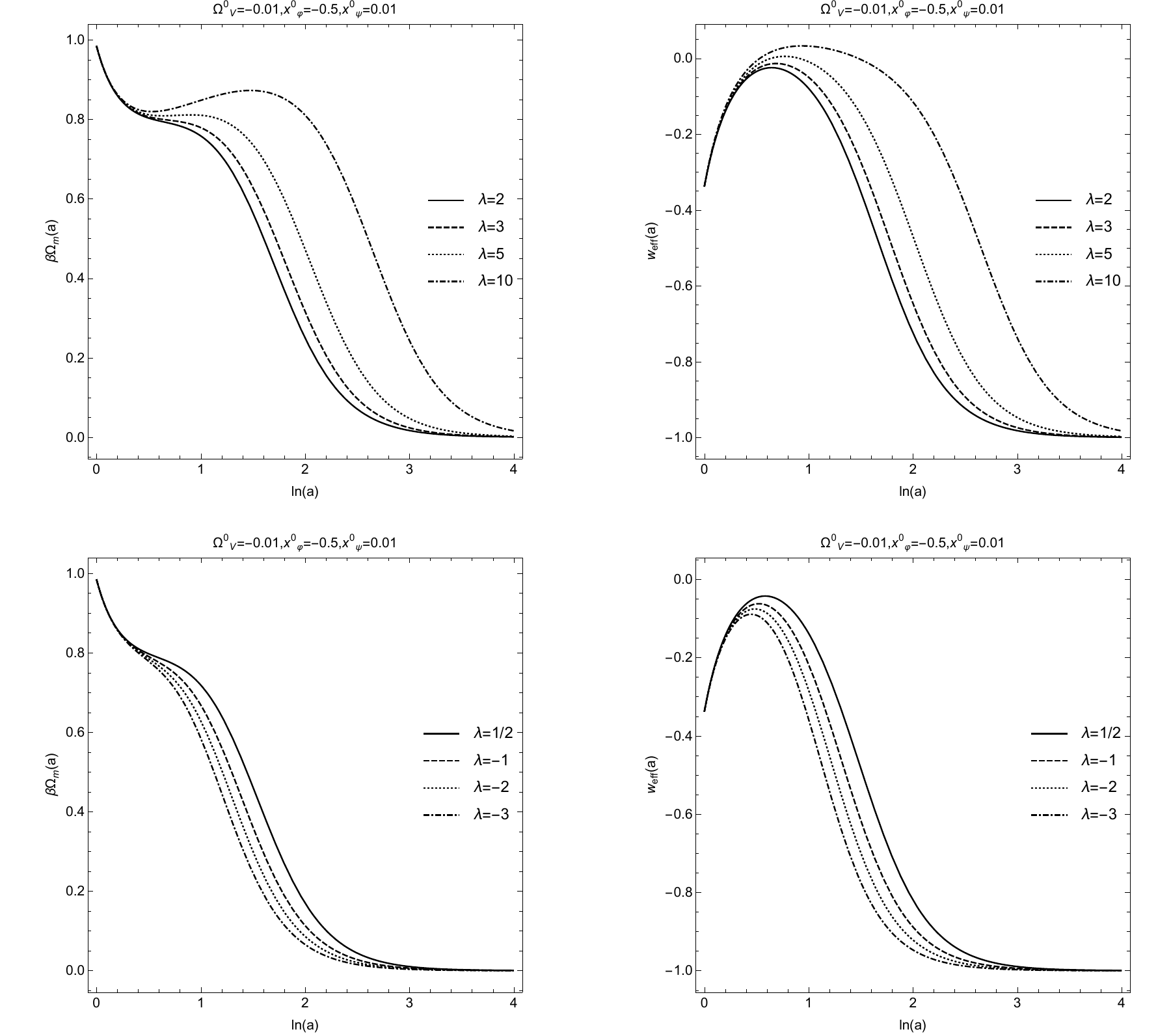}\caption{Qualitative
evolution of the physical parameters $\beta\Omega_{m}$ and $w_{eff}$ from
numerical similations of the dynamical (\ref{systA}) for the power-law
potential and for values of the free parameter $\lambda.$}%
\label{f1}%
\end{figure}

\begin{figure}[ptbh]
\centering\includegraphics[width=1\textwidth]{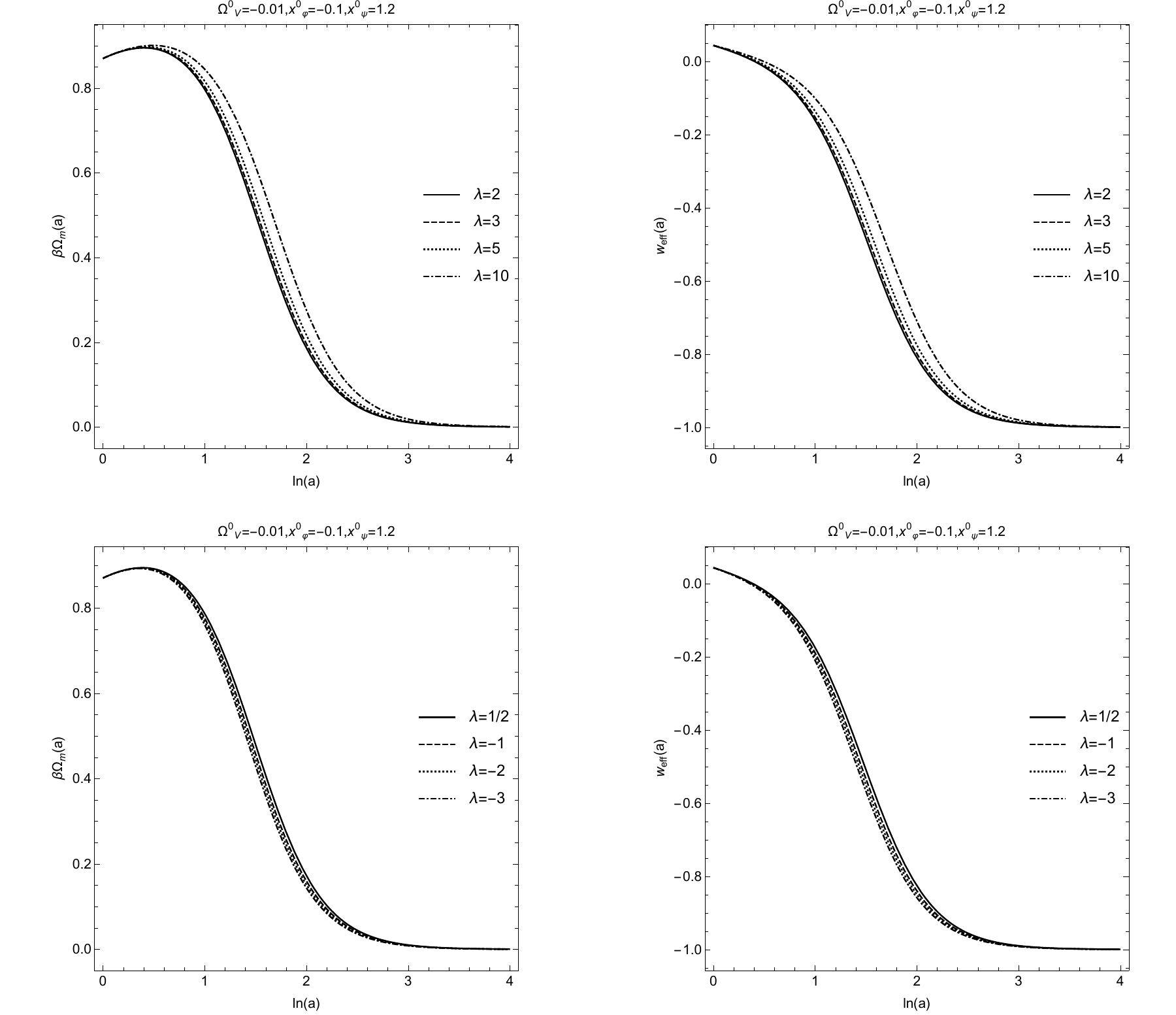}\caption{Qualitative
evolution of the physical parameters $\beta\Omega_{m}$ and $w_{eff}$ from
numerical similations of the dynamical (\ref{systA}) for the power-law
potential and for values of the free parameter $\lambda.$}%
\label{f2}%
\end{figure}

The stability properties are summarized in Table \ref{point2}. In Figs.
\ref{f1} and \ref{f2} we present the evolution of the physical parameter
$\beta\Omega_{m}$~and $w_{eff}$ given by numerical simulations for the
dynamical system (\ref{systA}) for the power-law potential and for different
set of initial conditions and values for the free parameter $\lambda$. The
initial conditions has been selected such that the trajectories to provide a
matter dominated epoch, point $B$ and reach the unique attract described by
point $D$. Moreover, we observe that parameter $\beta\Omega_{m}\geq0$, thus
the coupling parameter $\beta$ should be positive, in order $\Omega_{m}\geq0$. \ 

\begin{figure}[ptbh]
\centering\includegraphics[width=1\textwidth]{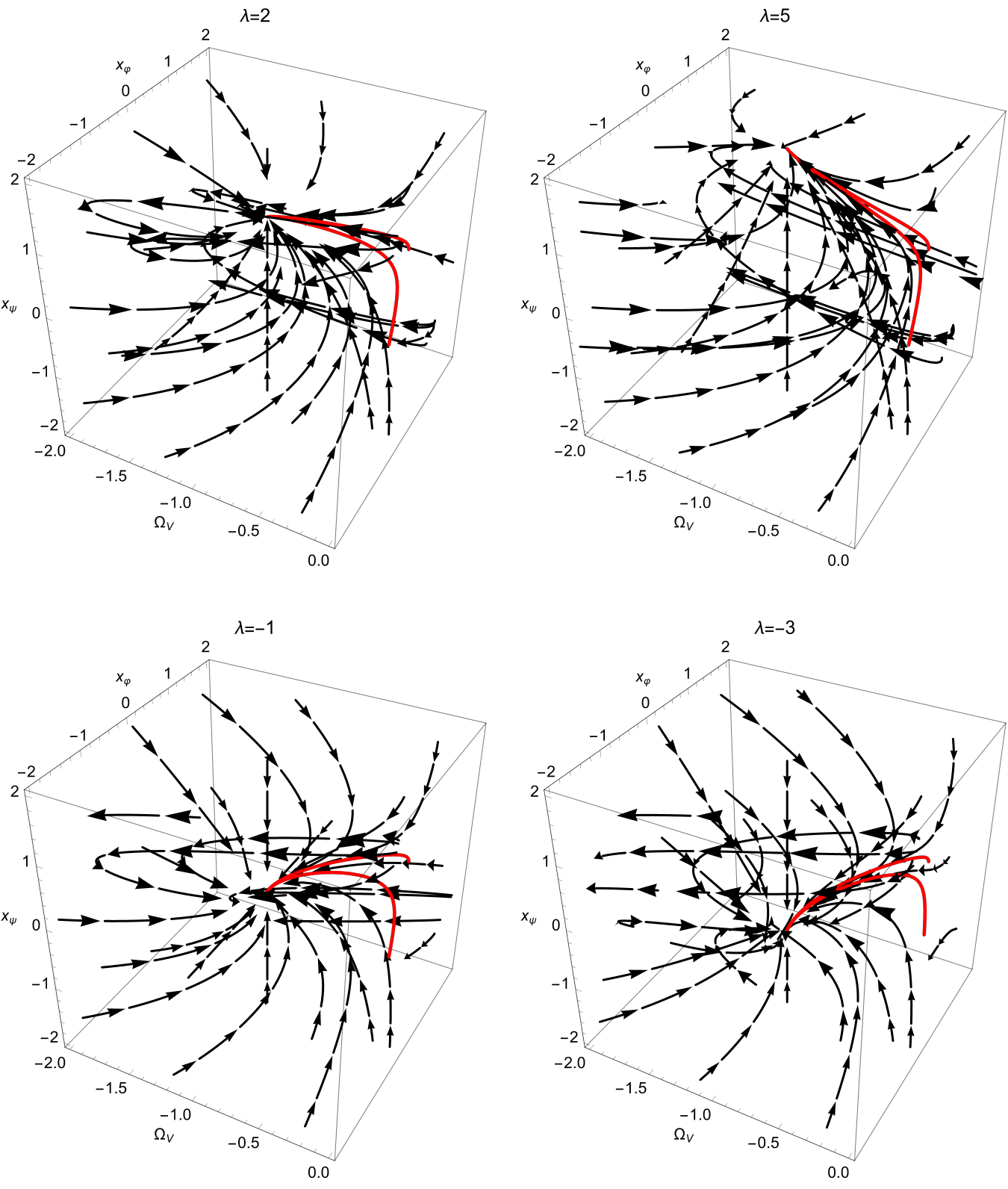}\caption{Phase-space
portraits for the dynamical system (\ref{systA}) with the power-law potential.
The red line corresponds to the initial conditions presented in Figs. \ref{f1}
and \ref{f2}. }%
\label{f3}%
\end{figure}

Moreover, in Fig. \ref{f3} we present the three-dimensional phase-space
portraits for the cosmological model with the power-law potential, where the
unique attractor is the de Sitter point $D$.%

\begin{table}[tbp] \centering
\caption{Stability properties for the power-law potential.}%
\begin{tabular}
[c]{cccc}\hline\hline
\textbf{Point} & \textbf{Eigenvalues} & \textbf{Type} & \textbf{Stability
Condition}\\\hline
$B$ & $3,-\frac{3}{2},-\frac{3}{2}$ & Matter dominated era & Saddle\\
$C$ & $0,3-\frac{2}{3x_{\psi}},\frac{6x_{\psi}-\lambda-1}{x_{\psi}}$ & Scaling
Solution & Attractor if $x_{\psi}<\frac{2}{3},\frac{6x_{\psi}-\lambda
-1}{x_{\psi}}<0$\\
$D$ & $\left\{  -3,-3,-3\right\}  $ & Potential-dominated &
Attractor\\\hline\hline
\end{tabular}
\label{point2}%
\end{table}%

\section{Conclusions}

\label{sec6}

In this study we examined the cosmological dynamics within the $f\left(
Q\right)  $ symmetric teleparallel gravity with a nonzero coupling between the
matter and gravity. For the background geometry we consider an isotropic and
homogeneous spatially flat FLRW geometry, while for the connection which
describes the gravitational field we select it to be defined in the
noncoincidence gauge. Within this consideration, dynamical degrees of freedom
are introduced by the connection within the field equations, leading to a
richer dynamical behaviour.

Although the gravitational theory is of fourth order, we select to work within
the scalar field description, where the field equations are expressed as
second-order equations with two scalar fields. One describes the dynamics of
the connection and the second scalar field attributes the higher-order
derivative components.

We employ the Hubble-normalization approach and we express the field equations
in terms of dimensionless variables. We define an equivalent system of
algebraic-differential equations and we express all the physical parameters in
terms of the new variables.

We investigate the phase space of the resulting dynamical system and we
explore the existence of stationary points. The stationary points identified
in the phase-space analysis represent distinct cosmological regimes, each
characterized by specific scalar field dynamics and energy contributions.
However, their physical relevance depends critically on their stability
properties. For a generic function $f\left(  Q\right)  $, we determined three
families of stationary points, which describe de Sitter solutions where the
potential term dominates, points $A$ and $D$; a matter-dominated epoch
described by point $B$, and scaling solutions given by the stationary points
$C$ and $E$.

For the power-law function $f\left(  Q\right)  \simeq Q^{\frac{\lambda
}{\lambda-1}}$, that is, the power potential $V\left(  \phi\right)  =V_{0}%
\phi^{\lambda}$, we perform a detailed analysis of the stability properties of
the stationary points. For this model, only the stationary points $B,$ $C$ and
$D$ exist. It follows that point $D$ is an attractor which can describe the
future acceleration of the universe. Point $B$ is a saddle point related to
the matter-dominated epoch, while the scaling solution described by point $C$
can be related to the early inflationary epoch.

By comparing these results with the previous study \cite{angrg} without the
interacting term, that is $\beta=0$, we observe that point $B$ is the new
stationary point which supports the matter-dominated era. Indeed, for
$\beta=0$, this stationary point is not supported. Consequently, for this
cosmological model, the introduction of the coupling function between the
scalar field and the matter source is essential for the description of the
matter epoch.

In future work we plan to investigate in detail the effects of the coupling
parameters within the cosmological perturbations, as well as to examine if
this cosmological model can describe the observable low-redshift expansion of
the universe.

\begin{acknowledgments}
AG was supported by Proyecto Fondecyt Regular 1240247. AP \& GL thanks the
support of VRIDT through Resoluci\'{o}n VRIDT No. 096/2022 and Resoluci\'{o}n
VRIDT No. 098/2022. Part of this study was supported by FONDECYT 1240514. The
authors thanks the Ionian University for the hospitality provided while this
work was carried out. Finally, the authors want to mention that this
collaboration began in the NEB-21 Conference, held in Corfu, which served as
the meeting point.
\end{acknowledgments}

\bigskip\textbf{Data Availability Statement:} Data sharing not applicable- no new data generated, the article describes entirely
theoretical research.

\end{document}